\documentclass[notitlepage, twocolumn, prl, superscriptaddress, floatfix]{revtex4-2}

\usepackage{tikz}
\usepackage{pgfplots}
\usetikzlibrary{intersections}
\pgfplotsset{compat=1.17}
\usepackage{mathtools}   
\usepackage{amsfonts}
\usepackage[normalem]{ulem}
\usepackage[flushleft]{threeparttable}
\usepackage{booktabs, makecell}
\usepackage{xcolor}
\usepackage{colortbl} 
\usepackage{placeins} 
\usepackage [english]{babel}
\usepackage{graphicx}
\usepackage{tikz}
\usepackage[final]{pdfpages}

\graphicspath{{figs/}}
\makeatletter
\AtBeginDocument{\let\LS@rot\@undefined}
\makeatother

\newcommand{\Wbar}{\bar{W}}
\newcommand{\Zbar}{\bar{Z}}

\begin{document}

\title{Phase chimera states: frozen patterns of disorder}
\author{Emma R.~Zajdela}
\email[]{\mbox{emmazajdela@u.northwestern.edu}}
\affiliation{Department of Engineering Sciences and Applied Mathematics, Northwestern University, Evanston, IL, USA}
\author{Daniel M.~Abrams}
\email[]{\mbox{dmabrams@northwestern.edu}}
\affiliation{Department of Engineering Sciences and Applied Mathematics, Northwestern University, Evanston, IL, USA}
\affiliation{Northwestern Institute for Complex Systems, Northwestern University, Evanston, IL, USA}
\affiliation{Department of Physics and Astronomy, Northwestern University, Evanston, IL, USA}

\begin{abstract}
Coupled oscillators can serve as a testbed for larger questions of pattern formation across many areas of science and engineering. Much effort has been dedicated to the Kuramoto model and phase oscillators, but less has focused on oscillators with variable amplitudes.  Here we examine the simplest such oscillators---Stuart-Landau oscillators---and attempt to elucidate some puzzling dynamics observed in simulation by us and others.  We demonstrate the existence and stability of a previously unreported state which we call a ``phase chimera state.'' Remarkably, in this state, the amplitudes of all oscillators are identical, but one subset of oscillators phase-locks while another subset remains incoherent in phase. We also show that this state can take the form of a ``multitailed phase chimera state'' where a single phase-synchronous cluster of oscillators coexists with multiple groups of phase-incoherent oscillators.  
\end{abstract}

\maketitle


Synchronization of coupled oscillators occurs across many natural systems, spanning biological, sociophysical, and engineered domains. Examples of such systems include the emergent synchronization of certain firefly species \cite{ermentrout1991}, pedestrian dynamics on the Millennium Bridge in London \cite{strogatz2005}, and the electric power grid's coordinated operation \cite{motter2013}. While synchronization can be beneficial in certain contexts, it can also contribute to detrimental effects in the human body, particularly in the context of neurological disorders such as epilepsy \cite{proix2018} and Parkinson's disease \cite{rivlin2006}. Therefore, understanding systems of coupled oscillators is important not only from a theoretical standpoint, but has implications for the wellbeing of social, natural, and infrastructure systems.

In the early 2000s, a surprising phenomenon was observed in the Kuramoto model, a simple model for coupled oscillators which only considers phase. The phenomenon consisted of synchronous and asynchronous subsets of oscillators coexisting in a stable configuration \cite{kuramoto2002}, despite all oscillators having identical natural frequencies and identical coupling to neighbors.  This was dubbed a ``chimera'' state \cite{abrams2004}, and has since garnered significant attention \cite{panaggio2015}. 

Although phase oscillator systems are themselves of considerable interest, there has also been a significant amount of work in  models which consider oscillators with non-constant amplitudes \cite{hakim1992, chabanol1997, ku2015, matthews1990, haugland2023}. In 2014, Sethia and Sen described a fascinating new discovery of \textit{amplitude mediated chimera states (AMC)} in the Stuart-Landau model \cite{sethia2014}. These particular states can exist for identical and globally coupled oscillators, which came as a major surprise---prior to their work, chimera states were thought to exist only in nonlocally coupled oscillators with phase-lagged coupling.  

The dynamics of the Stuart-Landau system of globally coupled identical oscillators are described by the following system of $N$ equations: 
\begin{equation}
\label{eq:SL}
    \dot{W_j} = W_j-(1+iC_2)|W_j|^2W_j+K(1+iC_1)(\Wbar-W_j) \;,
\end{equation}
$j=1 \ldots N$, where $W_j(t) \in \mathbb{C}$, $C_1, C_2 \in \mathbb{R}$, $K \in [0,1]$, $\Wbar = \left<W_j\right> \equiv N^{-1}\sum_{j=1}^N{W_j}$, and the ``overdot'' indicates a derivative with respect to time. 

\begin{figure}[t!]
\centering
\includegraphics[width=\linewidth]{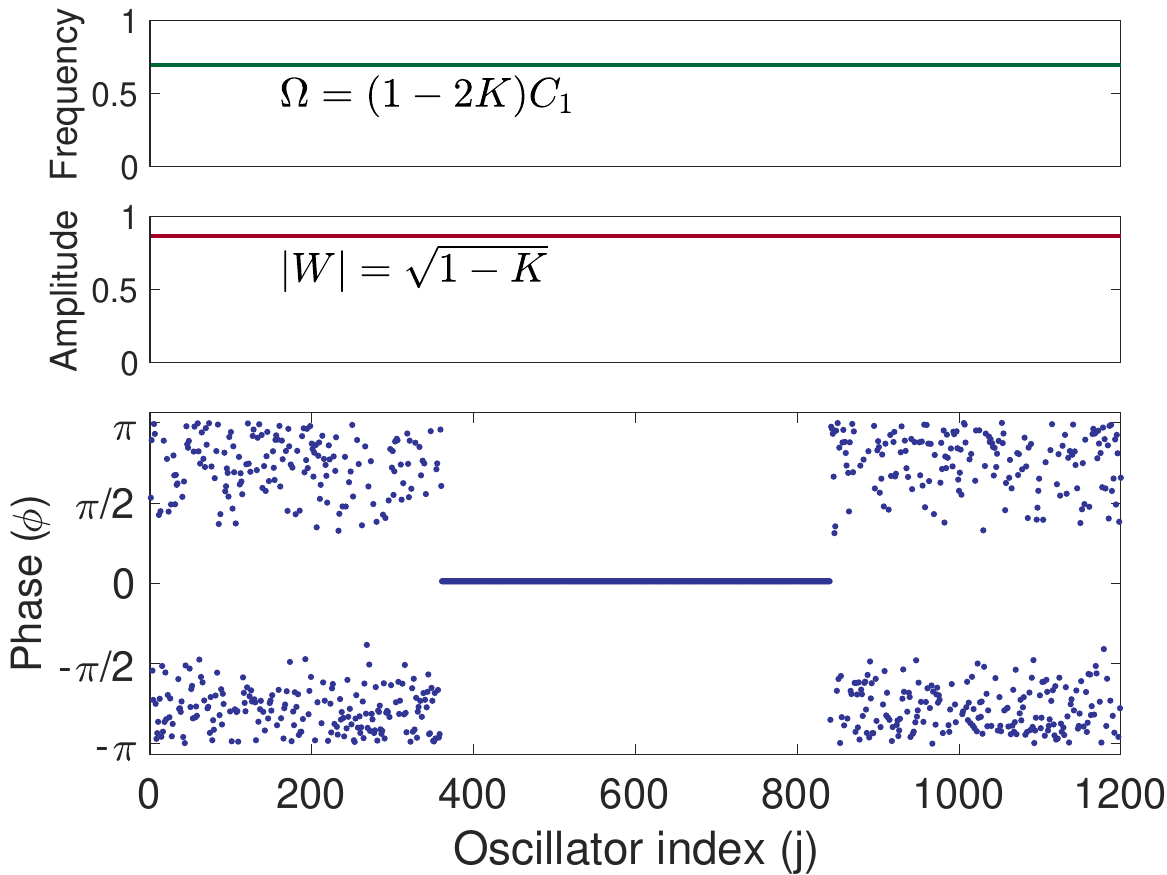}
\caption{\textbf{Phase chimera.} Top: long-term average of the frequencies $\dot{\phi}$; middle: snapshot of the amplitude profile $|W|$; bottom: snapshot of profile $\phi$.  $N=1200$ oscillators. Parameter values are $K = 0.25$, $x = 0.4$, $C_1 = 1.4$ and $C_2=-C_1$. The initial conditions consist of a two-cluster state with (nearly) uniform amplitude of $\sqrt{1-K}$ and phases centered on either $0$ and $\pi$ (each oscillator is perturbed by a random value from $\mathcal{N}(0, 0.01)$).
} 
\label{fig:phaseChimera}
\end{figure}

\begin{figure*}[thb!]
\includegraphics[width=\linewidth]{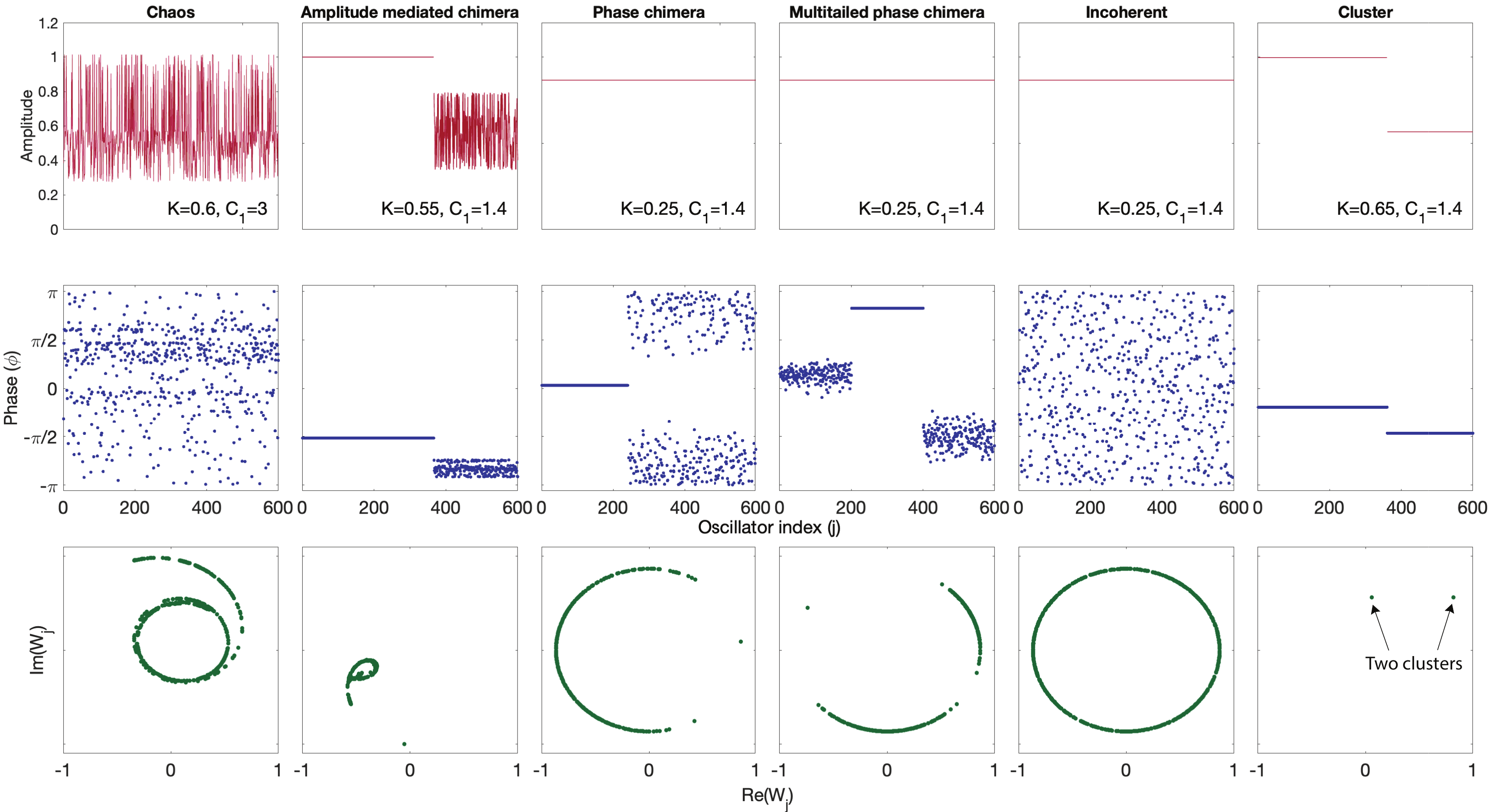}
\caption{\textbf{Zoology of states in the Stuart-Landau system.} Amplitude (top row), phase (middle row), and distributions in the complex plane (bottom row) for various states observed in the Stuart-Landau system of coupled oscillators. Here, $N=600$ and $C_2=-C_1$. For all but the chaotic state, $C_1=1.4$. For the chaotic state, $C_1=3$ and $K=0.6$ with initial condition (IC) randomly distributed on the unit disk. For the amplitude mediated chimera (AMC), $K=0.55$ and IC is a two-cluster state ($x=0.4$ with amplitude $\sqrt{1-K}$ and phases at $0$ and $\pi$). For the phase chimera, multitailed phase chimera, and incoherent states $K=0.25$. ICs for these three states are, respectively: a two-cluster state ($x=0.4$ with amplitude $\sqrt{1-K}$ and phases at $0$ and $\pi$); a three cluster state ($1/3$ of oscillators in each cluster) with amplitudes for two of the clusters $\sqrt{1-K}$ and phases $\pi$ and $0$, and the third cluster also at phase $0$ at amplitude $0.97\sqrt{1-K}$ (though many choices work similarly); uniform randomly distributed oscillators on the unit disk. Note that supplemental videos accompanying this paper also illustrate several of these states.}
\label{fig:states}
\end{figure*}

In this paper, we report our discovery of a remarkable new species in the zoology of chimera states: the \textbf{\textit{phase chimera}} (see Fig.~\ref{fig:phaseChimera}). This state is characterized by all oscillators having the same amplitude and frequency values. One subset, however, has identical locked phases, while the remaining oscillators exhibit a distribution of phases. We analytically demonstrate the existence of this state and find proof of its stability in certain regions of the parameter space.

We also report another related new type of state, which we call the \textit{multitailed phase chimera} in analogy to multiheaded chimera states which have been previously observed \cite{larger2013,maistrenko2014}. In multiheaded chimera states, multiple clusters of coherent oscillators coexist with incoherent oscillators whereas in multitailed chimera states, one cluster of coherent oscillators coexists with multiple defined regions of incoherent oscillators. 

\textbf{Numerical illustration}---We originally discovered the phase chimera state while investigating the stability of the two-cluster state with fraction $x$ of oscillators in one cluster and $1-x$ in the other, under the simplifying assumption that $C_2=-C_1$. In numerical simulations, we perturbed the amplitude of each oscillator by a small normally distributed amount and noticed that, for certain parameter values, the two-cluster state converged to this surprising new state. In this state, illustrated in Fig.~(\ref{fig:phaseChimera}), the amplitudes of the oscillators have a fixed universal value of $A = \sqrt{1-K}$, the phases of fraction $x$ of the oscillators are identical and locked, the phases of fraction $1-x$ oscillators are incoherent, and the long-term average frequency of all oscillators is uniform. We call this new state a \textit{\textbf{``phase chimera.''}} 

\begin{figure}[t!]
\includegraphics[width=0.7\linewidth]{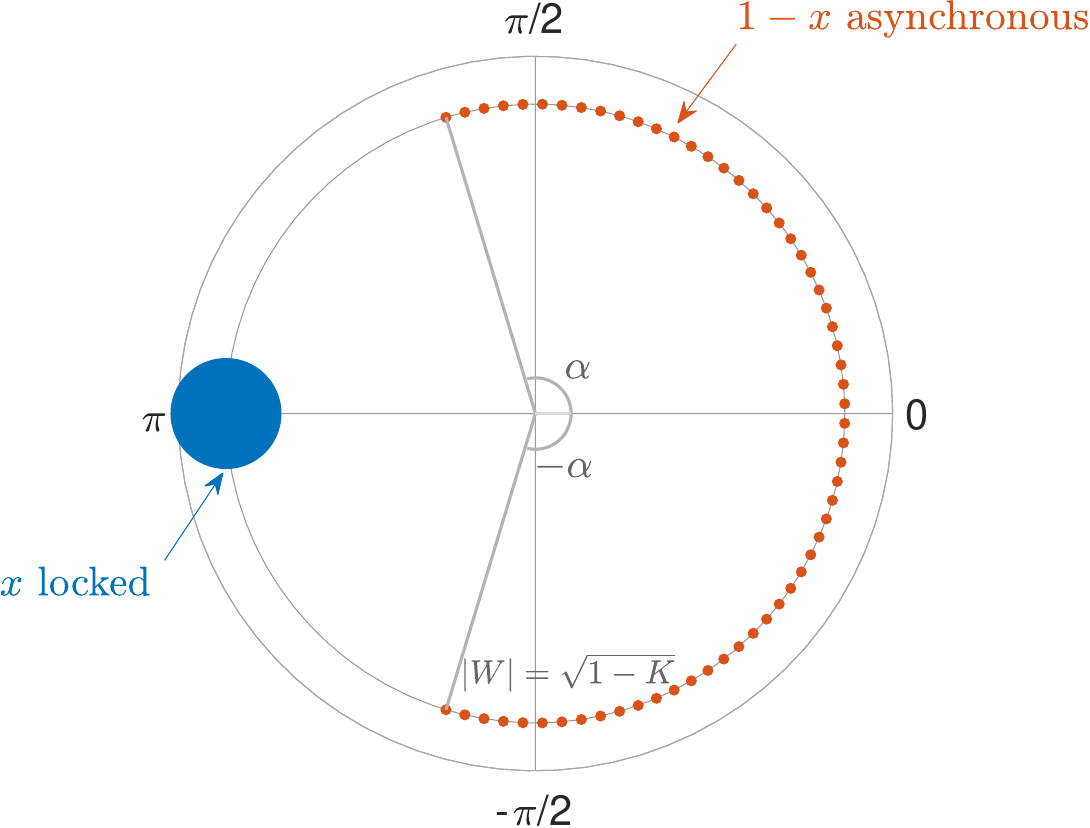}
\caption{\textbf{Phase chimera and $x$-$\alpha$ relationship.} Location on the unit disk of $N$ oscillators with fraction $x$ in the locked group (larger blue dot) and $1-x$ in the incoherent group (set of smaller red dots). All oscillators are located at amplitude $|W|=\sqrt{1-K}$. The oscillators in the incoherent group are uniformly distributed from $-\alpha$ to $\alpha$.  
}
\label{fig:xalpha}
\vspace{-0.05cm}
\end{figure}

Although many different distributions of phase-incoherent oscillators can exist in the phase chimera, we choose for simplicity to study a version where incoherent oscillators are distributed symmetrically about a point separated by $\pi$ radians from the locked oscillators (all are located on the circle of radius $\sqrt{1-K}$).  In our analytical work (see below), we consider a rotating frame in which the locked oscillators are located on the negative real axis, and the incoherent oscillators are uniformly distributed between angles $-\alpha$ and $\alpha$. See Fig.~(\ref{fig:xalpha}) for a graphical display of this. The incoherent portion of the chimera state behaves similarly to the splay \cite{strogatz1993} and fully incoherent states (see Fig.(~\ref{fig:states}), with the difference that those states include oscillators with phases covering the full domain from $[-\pi,\pi]$. 

In Fig.~\ref{fig:states}, we show various stable equilibrium states for the globally-coupled Stuart-Landau system under different initial conditions and parameter values. Other stable equilibrium states are possible but not shown (e.g., fully synchronous state, three-cluster state, splay states---where oscillators are uniformly distributed on the circle of radius $\sqrt{1-K}$). Note that the phase chimera, multitailed phase chimera, and incoherent states can all stably coexist with the same parameter values and are initial-condition dependent. 

\textbf{Analytical demonstration of existence}---Here, we examine requirements for existence of the phase chimera state. To study the system in a co-rotating frame with frequency $\Omega$, we let $W(t) = Z(t)e^{i\Omega t}$. We then rewrite Eq.~(\ref{eq:SL}) as
\begin{equation}
\label{eq:SL_corot}
    \dot{Z_j} = (1-i\Omega)Z_j-(1+iC_2)|Z_j|^2Z_j+K(1+iC_1)(\Zbar-Z_j).
\end{equation}

Assume in the co-rotating frame that each oscillator has phase $\theta_j$ and fixed universal amplitude $A$, so $Z_j=Ae^{i\theta_j}$. We define the centroid for the system as $\Zbar = \left<Z_j\right> = A \left<e^{i\theta_j}\right> = ARe^{i\Psi}$, where $R \in [0,1]$ is the (real) order parameter for the system and $\Psi$ the mean phase.

Substituting this into Eq.~(\ref{eq:SL_corot}) and taking real and imaginary parts, we obtain the following system of equations: 
\begin{equation} \label{eq:SL4a}
    \dot{\theta_j} = KR[C_1\cos(\Psi-\theta_j ) + \sin(\Psi-\theta_j ) ] -A^2C_2 - KC_1 - \Omega 
\end{equation}
\begin{equation} \label{eq:SL4b}
    0 = KR[C_1\sin(\Psi-\theta_j )-\cos(\Psi-\theta_j )] + A^2 + K - 1.    
\end{equation}


Equation (\ref{eq:SL4b}) has no solution other than the fully phase-locked state unless $R=0$, meaning that the centroid of the system must be $0$ for any more ``interesting'' state to exist. This condition further implies that $A=\sqrt{1-K}$ for all oscillators. Substituting into Eq.(\ref{eq:SL4a}), we find that all oscillators must be frequency locked at frequency $\Omega = K(C_2-C_1)-C_2$ and are stationary in the co-rotating frame. These theoretical values of $A$ and $\Omega$ are consistent with our numerical observations. The fully incoherent, splay, phase chimera, and multitailed phase chimera states all satisfy this ``balance equation'' $R=0$. 

In the case of the phase chimera, the condition $R=0$ implies a relationship between the fraction of locked oscillators $x$ and the distribution of incoherent oscillators.  Locating the locked oscillators at $\theta=\pi$ (without loss of generality),
\begin{align}  \label{eq:xalphaa}
    R = 0 &= \left<e^{i \theta_j}\right> = \left<e^{i \pi}\right>_{\textrm{locked}} + \left<e^{i \theta_j}\right>_{\textrm{inc.}} \\
     &= -x + \frac{1}{N}\sum_{j \in \textrm{\;inc.~group}} e^{i\theta_j}\;.  \notag
\end{align}
Assuming a uniform distribution of angles for incoherent oscillators (see Fig.~(\ref{fig:xalpha})), we can relate fraction $x$ to the maximum angle $\alpha$ as
\begin{equation} 
    x = \frac{1}{N}\frac{\sin(\frac{\alpha(1-x)N}{Nx - N + 1})}{\sin(\frac{\alpha}{Nx - N + 1})} \xrightarrow[N \to \infty]{} \frac{\sin(\alpha)}{\sin(\alpha)+\alpha}. 
\end{equation}

%


\textbf{Stability analysis}---We now examine the stability boundary of the phase chimera state. We make the simplifying assumption that $C_2=-C_1$, however, these method can be generalized to the case where $C_2\neq C_1$ \footnote{See Supplemental Material (SM) at [URL will be inserted by publisher] for additional detail.}

\begin{figure}[b!]
    \centering
    \begin{tikzpicture}
        \node[anchor=south west,inner sep=0] (main) at (0,0) {\includegraphics[width=\linewidth]{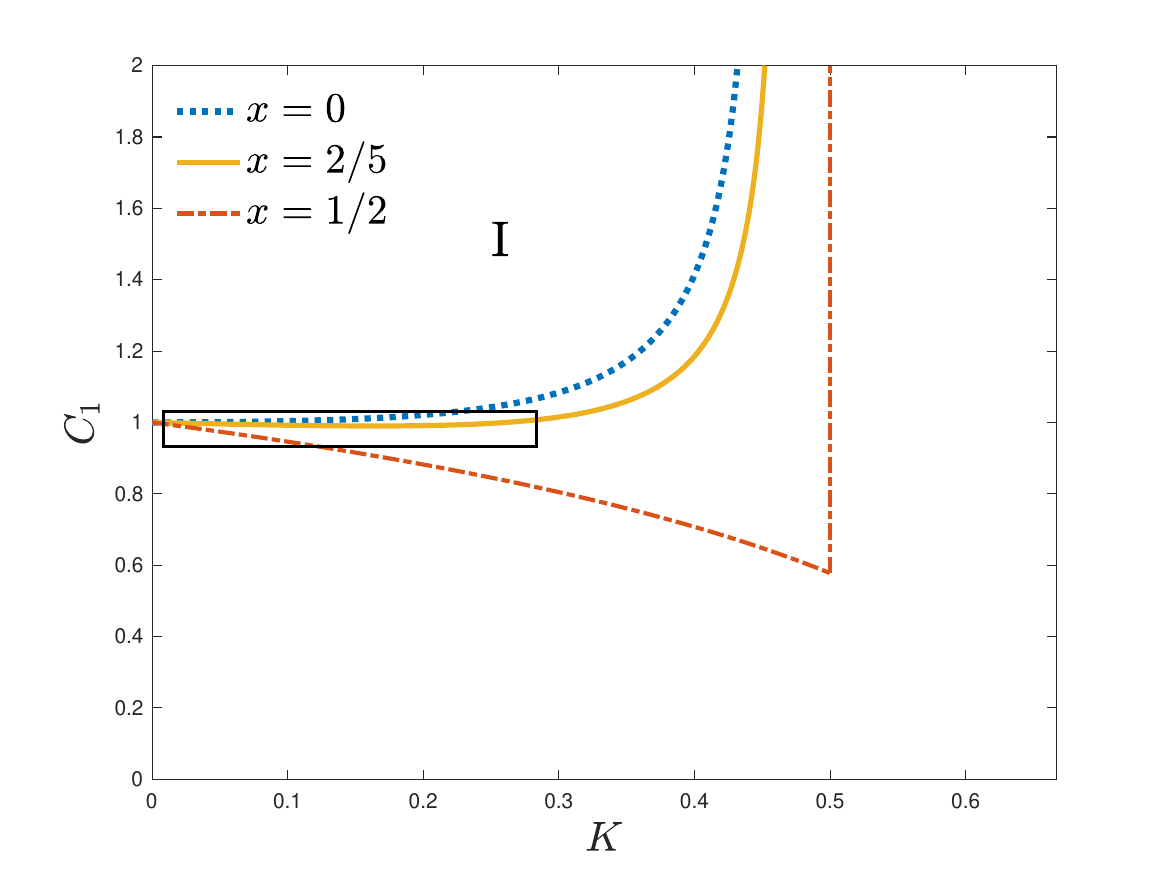}};
        \begin{scope}[x={(main.south east)},y={(main.north west)}]
            \node[anchor=south west] at (0.15,0.12) {\includegraphics[width=.3\linewidth]{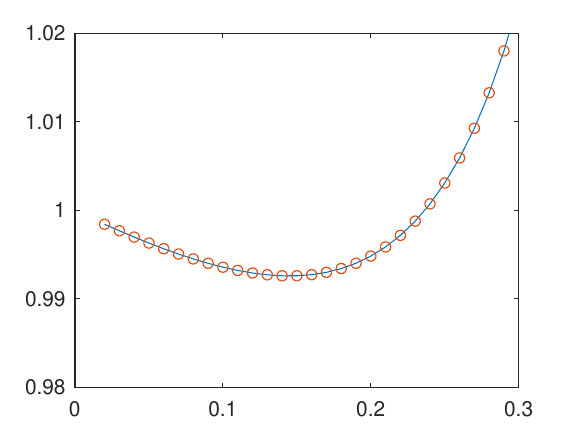}};
        \end{scope}
    \end{tikzpicture}
    \caption{\textbf{Stability boundaries.} Blue dotted, yellow solid, and red dot-dashed curves show stability boundaries for the phase chimera with varying fractions of locked oscillators $x = 0$, $x=2/5$, and $x=1/2$, respectively, and $C_2=-C_1$. The phase chimera is stable in region I above the curves and unstable elsewhere. $x=0$ case is equivalent to the stability curve for the fully incoherent state. Inset shows a zoomed version of the stability curve for $x=2/5$ and demonstrates agreement between analytical results (blue solid curve) and numerical computation of eigenvalues for system with $N=5$.}
    \label{fig:stability}
    \vspace{-0cm}
\end{figure}

\begin{figure*}[ht!]
\centering
\includegraphics[width=0.8\linewidth]{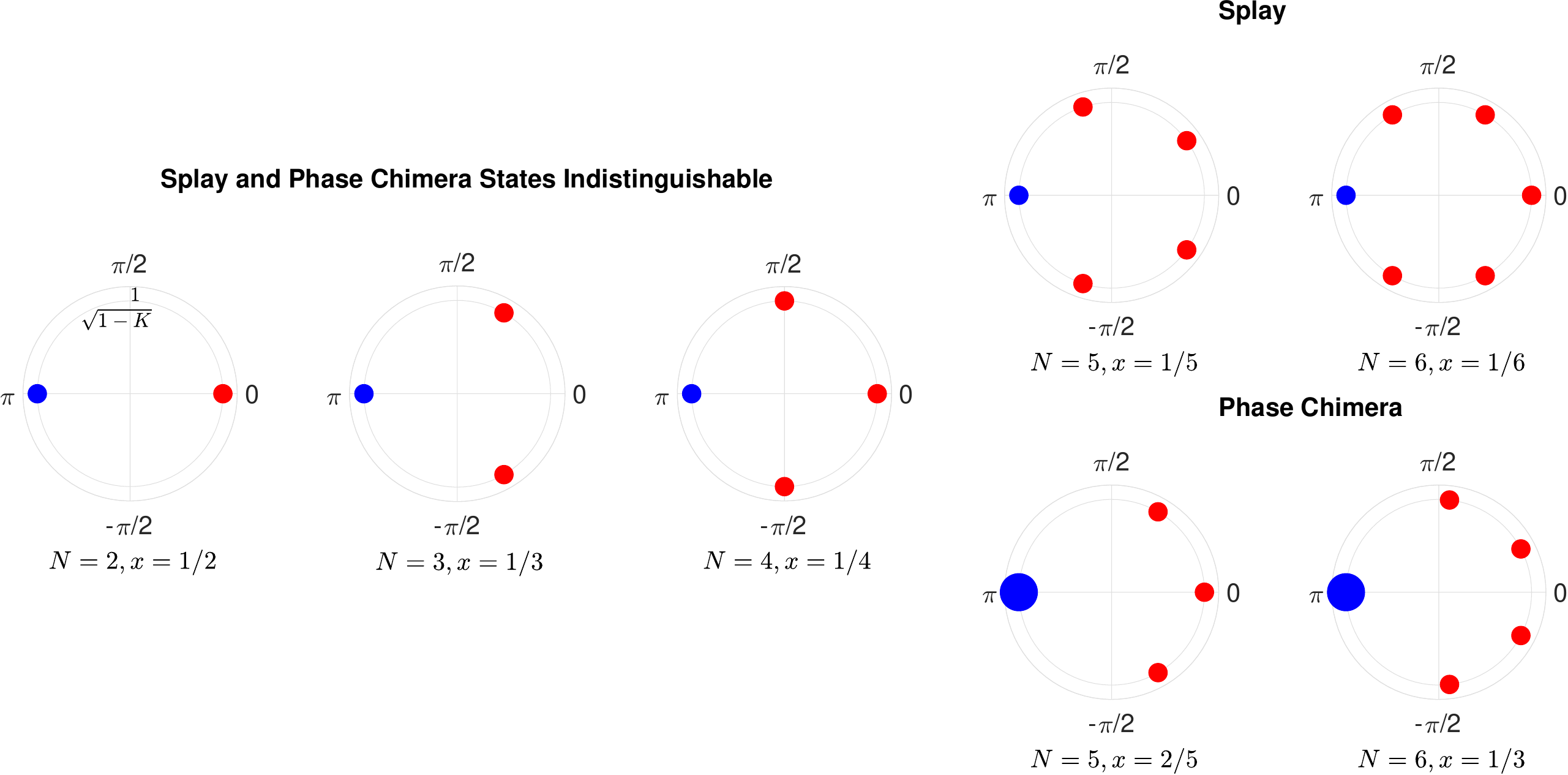}
\caption{\textbf{Minimal chimera.} Location on the unit disk of the oscillators for small $N$ values ($N=2,3,4,5,6$). Parameter values are chosen in the range where the phase chimera and splay states are stable with values $K = 0.25$, $C_1 = 1.4$ and $C_2=-C_1$. In each panel a fraction $x$ of oscillators is locked at $\theta =\pi$ (blue dots) and a fraction $1-x$ of oscillators are distributed uniformly about $\theta=0$ (red dots); all amplitudes are $|W|=\sqrt{1-K}$. Oscillators in blue (located at $\theta=\pi$).  Dot sizes are proportional to the number of oscillators in the group.} 
\label{fig:smallN}
\end{figure*}

We use a method similar to that used for determining the stability boundary of the incoherent state in references \cite{hakim1992} and \cite{chabanol1997}. We peturb the amplitudes of oscillators in the phase chimera state by an amount $\epsilon_j$, setting 
\begin{equation}
    Z_j = \left(\sqrt{1-K}+\epsilon_j\right)e^{i\theta_j}
\end{equation}
in Eq.~\eqref{eq:SL_corot} and linearizing to obtain 
\begin{equation*}
    \dot{\epsilon_j} = (1+iC_1)\frac{K}{N}\sum_{k=1}^N \epsilon_k e^{i(\theta_k-\theta_j)}-(1-iC_1)(1-K)(\epsilon_j+\epsilon_j^*) \;,
\end{equation*}
where $\epsilon_j^*$ denotes the complex conjugate of $\epsilon_j$ (see SM for details).

The system linearized about the phase chimera state is of dimension $2N\times2N$ and has $N-2$ negative eigenvalues equal to $-2(1-K)$ and $N-2$ zero eigenvalues corresponsing to neutral perturbations that preserve the centroid.  This leaves a distinguished four dimensional subspace that we examine through the change of variables $\epsilon_j=a e^{-i\theta_j} + b e^{i\theta_j}$. The dynamics of $(a,b^*,b,a^*)$ are governed by the $4\times 4$ matrix:
\begin{equation}
\begin{pmatrix}
u & v & w & 0\\
v^* & v^* & 0 & 0\\
0 & 0 & v & v\\
0 & w^* & v^* & u^*
\end{pmatrix},
\end{equation}
where $u=2K-1+iC_1$, $v = -(1-iC_1)(1-K)$, and $w=K(1+iC_1)\delta$, where $\delta = \left<e^{2i\theta_k} \right> \in [0,1]$ (see SM for additional detail). For the phase chimera state described above, 
\begin{equation}
\scalebox{0.83}{$\delta = x\left\{1+\frac{\cos[\frac{\alpha(1-x)N}{Nx - N + 1}]}{\cos[\frac{\alpha}{Nx - N + 1}]}\right \} \xrightarrow[N \to \infty]{} x\left \{\cos[\textrm{arcsinc}(\frac{x}{1-x})]+1\right \} \;.$} 
\end{equation}
We apply the Routh-Hurwitz criterion (see, e.g., \cite{gantmacher1992}) to this matrix and find that $K < 1/2$ is required for stability of the phase chimera state, as well as the more complicated condition
%
%
%

\scalebox{0.88}{
$\begin{aligned}
C1^2>& \frac{-9(K - 2/3)}{(9\delta^2 - 7)K^4 + (12-21\delta^2)K^3 + 4[(4\delta^2 - 1)K^2 - \delta^2 K]} \notag \\&\phantom{12}\times \{(-4/3K^2 + 2K - 2/3)\Delta \notag\\ &\phantom{1234} + [(\delta^2 - 3)K^2 + (-\delta^2 + 6)K - 2](K - 2/3)\}
\end{aligned}$}


where $\Delta = \sqrt{4 + (-8\delta^2 + 9)K^2 + (8\delta^2 - 12)K}$.

\vspace{5mm}
When $x \to 0^+$ (limiting to the fully incoherent and splay cases), this simplifies to: 
\begin{equation}
    C_1 > \frac{2-3K}{\sqrt{7K^2-12K+4}} \;.
\end{equation}

For $x>0$, the region of (neutral) stability of the phase chimera is actually larger than the corresponding region for the splay and incoherent states (see Fig.~(\ref{fig:stability})). The region is largest for $x \to 1/2^{-}$, when the system becomes degenerate with the symmetric two-cluster state (separated by $\pi$) and $\delta \to 1$ is maximal.  Then the boundary is given by 
\begin{equation}
     C_1 > \sqrt{\frac{2-3K}{2-K}}.
\end{equation}

\textbf{Minimal phase chimera}---A ``degenerate'' phase chimera state exists and can be observed for $N$ as small as $N=2$ (see Fig.~\ref{fig:smallN}); in that case, the phase chimera, two-cluster state and splay states are indistinguishable with phases separated by $\pi$. For $N=3$ and $N=4$, the phase chimera exists for $x=1/N$ and is equivalent to the splay state. However, for $N\geq5$, the phase chimera is different from the splay state. Fig.~\ref{fig:smallN} shows examples for $N=5$ and $N=6$, when the phase chimera and splay states become distinguishable for $x=2/N$. 

We believe that these minimal phase chimera states could allow for further investigation into the origin of chimera states.  In particular, we hypothesize that they are born from clustered states (namely the two-cluster state), as has been proposed in prior work \cite{schmidt2015} in systems with more complicated coupling \cite{haugland2023}. 

\textbf{Multitailed phase chimera}---Chabanol, Hakim, and Rappel aptly noted in 1997, ``There are many different ways to realize condition [Eq.~(\ref{eq:xalphaa})].'' Among the ways to realize this condition are the incoherent and splay states which were previously reported and our newly discovered \textit{phase chimera state}. However, other interesting states also satisfy this initial condition which we call \textbf{\textit{multitailed phase chimera states}}. They are analogous to multiheaded chimera states \cite{larger2013,maistrenko2014} (also known as multicomponent \cite{ujjwal2013}, multicluster \cite{xie2014,yao2015}, or multiple-headed chimera \cite{larger2015,schmidt2017}), which consist of multiple clusters of coherent oscillators among incoherent oscillators. Since we observe a state consisting of one locked cluster of oscillators among multiple incoherent clusters, we name it a \textit{multitailed phase chimera} state. See Fig.~\ref{fig:states} for a graphical representation of this state in the case where there is one cluster of locked oscillators and two groups of incoherent oscillators, which was obtained from the perturbation of an initial condition consisting of three clusters. 

\textbf{Conclusions}---The findings presented in this study unveil a new class of chimera state: the phase chimera, which (unlike other chimera states) is frozen in a rotating frame of reference. The phase chimera can come in many guises, depending on the choice of initial conditions defining the distribution of the incoherent oscillators. Perhaps surprisingly, this state has a larger domain of existence and (neutral) stability than incoherent or splay states, raising the question of why it hasn't previously been noticed. 


This discovery opens pathways for subsequent research to address some of the immediate questions raised: e.g., what are the origins of the phase chimera---does it emerge from a bifurcation off of a clustered state? Why does it appear so readily in simulations, despite the neutral stability?  Can we define a basin of attraction for this state within some restricted manifold?  And, of course, do states such as this emerge in naturally occuring systems?

\vspace{5mm}

\begin{acknowledgments}
The authors would like to thank the National Science Foundation Graduate Research Fellowship Program DGE-184216 for financial support.
\end{acknowledgments}

\newpage

%

\clearpage



\section*{Supplementary material}

# Supplemental material for
# Phase chimera states: frozen patterns of disorder

Emma R. Zajdela[1, *] and Daniel M. Abrams[1, 2, 3, †]

[1]*Department of Engineering Sciences and Applied Mathematics, Northwestern University, Evanston, IL, USA*
[2]*Northwestern Institute for Complex Systems, Northwestern University, Evanston, IL, USA*
[3]*Department of Physics and Astronomy, Northwestern University, Evanston, IL, USA*

### DETERMINING THE PHASE CHIMERA STABILITY BOUNDARY

This section follows the method outlined in chapter 2 of the 1997 Ph.D. thesis by Chabanol [1, 2], where she determines the stability boundary of what she calls the asynchronous regime (i.e., the regime where oscillators are moving around a circle with center of mass 0 on average). Recall from the ***Analytical demonstration of existence*** section of the main text that, in a co-rotating frame with frequency $\Omega = K(C_2 - C_1) - C_2$, the phase chimera state can be written as

$$Z_j = \sqrt{1-K} e^{i\theta_j} , \tag{1}$$

where the phases $\theta_j$ satisfy the balance condition

$$\sum_{j=1}^{N} e^{i\theta_j} = 0 . \tag{2}$$

A perturbation to the phase chimera state in the co-rotating frame is introduced with the form

$$Z_j = (\sqrt{1-K} + \epsilon_j)e^{i\theta_j} , \tag{3}$$

where $\epsilon_j$ is a complex quantity. Substituting equation Eq. (3) into the Stuart-Landau equation in the co-rotating frame,

$$\dot{Z}_j = (1-i\Omega)Z_j - (1+iC_2)|Z_j|^2 Z_j + K(1+iC_1)(\bar{Z} - Z_j) , \tag{4}$$

where $\bar{Z} = \frac{1}{N}\sum_{k=1}^{N} Z_k$, gives us the following equation:

$$\dot{\epsilon}_j = \frac{K(1+iC_1)}{N}\sum_k \epsilon_k e^{i(\theta_k-\theta_j)} - (1+iC_2)(\sqrt{1-K}+\epsilon_j)(|\sqrt{1-K}+\epsilon_j|^2 + K - 1) . \tag{5}$$

After linearization, we obtain the following equation for the perturbation $\epsilon_j$, with $\epsilon_j^*$ its complex conjugate:

$$\dot{\epsilon}_j = \frac{K(1+iC_1)}{N}\sum_k \epsilon_k e^{i(\theta_k-\theta_j)} - (1+iC_2)(1-K)(\epsilon_j + \epsilon_j^*). \tag{6}$$

In order to determine the eigenvalues of the system (and hence the phase chimera state's stability), we follow Chabanol's method [1, 2] and note that the coupling term $\frac{K(1+iC_1)}{N}\sum_k \epsilon_k e^{i(\theta_k-\theta_j)}$ is contained in a four dimensional subspace $E_4$ with basis vectors $\{\epsilon_j^{(1)} = \cos(\theta_j), \epsilon_j^{(2)} = \sin(\theta_j), \epsilon_j^{(3)} = i\cos(\theta_j), \epsilon_j^{(4)} = i\sin(\theta_j)\}$. The orthogonal complement of the subspace $E_4$ is contained in the kernel of the linear operator defined by the coupling term. Furthermore, $E_4$ is invariant under the second part of the operator ($\epsilon_j \to -(1+iC_2)(1-K)[2\Re(\epsilon_j)]$). Therefore, we can determine the eigenvalues of the system on $E_4$ and on its orthogonal complement individually.

Let's start by considering the orthogonal complement of $E_4$, which has a dimension of $2N - 4$. In the case where ($\epsilon_j$) is orthogonal to $E_4$, the coupling term is null, so the dynamics of the system simplify to

$$\dot{\epsilon}_j = -(1+iC_2)(1-K)(\epsilon_j + \epsilon_j^*) . \tag{7}$$

Eigenvalues for this complex ODE can be computed by various methods. Here we express it in terms of real and imaginary parts of $\epsilon_j$:

$$\begin{pmatrix} -2(1-K) & 0 \\ -2C_2(1-K) & 0 \end{pmatrix}. \tag{8}$$

Since this has lower triangular form, we can now read off the two distinct eigenvalues, each of which is repeated $N-2$ times.

The zero eigenvalues correspond to eigenvectors of the type $\epsilon_j = ix_j$, where $x_j \in \mathbb{R}$ satisfies the centroid-preserving constraints $\sum_j x_j \cos(\theta_j) = 0$ and $\sum_j x_j \sin(\theta_j) = 0$. The eigenvalues $-2(1-K)$ have associated eigenvectors $\epsilon_j = (1+C_2)y_j$, where $y_j \in \mathbb{R}$ satisfies the same centroid-preserving constraints as $x_j$. Therefore, within this subspace, the system is not unstable as long as coupling strength $K < 1$ (we focus anyway on coupling strengths $0 < K < 1$).

On the subspace $E_4$, we can write the perturbation in the form $\epsilon_j = ae^{-i\theta_j} + be^{i\theta_j}$ where $a$ and $b$ are complex numbers. Then, substituting into Eq. (6), we obtain:

$$\dot{a}e^{-i\theta_j} + \dot{b}e^{i\theta_j} = \frac{K(1+iC_1)}{N} e^{-i\theta_j} \sum_k \left(a + be^{2i\theta_k}\right) - (1+iC_2)(1-K)\left((a+b^*)e^{-i\theta_j} + (a^*+b)e^{i\theta_j}\right). \quad (9)$$

This equation must hold for any phase $\theta_j$ which satisfies the balance condition Eq. (2), so for convenience we substitute $\theta_j = 0$ and $\theta_j = \frac{\pi}{2}$ into Eq. (9) and then evaluate the sum and difference of the two resulting equations to obtain:

$$\dot{a} = bK(1+iC_1)\frac{1}{N}\sum_k e^{2i\theta_k} + aK(1+iC_1) - (1+iC_2)(1-K)(a+b^*), \quad (10)$$

$$\dot{b} = -(1+iC_2)(1-K)(a^*+b). \quad (11)$$

The dynamics of the system in the basis $(a, b^*, b, a^*)$ is thus given by the $4\times 4$ matrix,

$$\begin{pmatrix} u & v & w & 0 \\ v^* & v^* & 0 & 0 \\ 0 & 0 & v & v \\ 0 & w^* & v^* & u^* \end{pmatrix}, \quad (12)$$

where $u = K(1+iC_1) - (1+iC_2)(1-K)$, $v = -(1+iC_2)(1-K)$, and $w = K(1+iC_1)\delta$ with $\delta = \frac{1}{N}\sum_{k=1}^N e^{2i\theta_k}$. This is the system we will apply the Routh-Hurwitz criterion to in order to establish (neutral) stability of the phase chimera state.

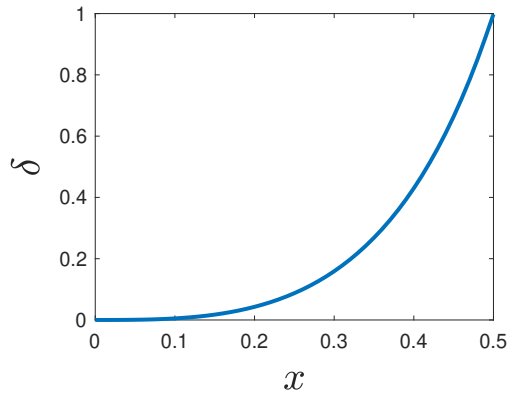

Figure 1. Relationship between the fraction $x$ of locked oscillators and the parameter $\delta = \left\langle e^{2i\theta_k}\right\rangle$ for $N=1000$. The parameter $\delta$ varies between 0 and 1. The parameter $x$ varies between 0 and $1/2$, which is the domain of existence of the phase chimera state.

For the phase chimera state, we wish to relate $\delta$ to $x$, but $\alpha$ cannot be readily eliminated from the equation enforcing that the centroid remain at zero:

$$x = \frac{1}{N} \frac{\sin(\frac{\alpha(1-x)N}{Nx-N+1})}{\sin(\frac{\alpha}{Nx-N+1})} \xrightarrow{N\to\infty} \frac{\sin(\alpha)}{\sin(\alpha) + \alpha} \ . \tag{13}$$

Therefore for finite $N$ we write $\delta$ in terms of both $\alpha$ and $x$ (with $\alpha$ solving the implicit equation above):

$$\delta = x \left\{ 1 + \frac{\cos[\frac{\alpha(1-x)N}{Nx-N+1}]}{\cos[\frac{\alpha}{Nx-N+1}]} \right\} \xrightarrow{N\to\infty} x \left\{ \cos[\text{arcsinc}(\frac{x}{1-x})] + 1 \right\} . \tag{14}$$

The implied function $\delta(x)$ is shown in Fig. (1) for large (but finite) $N$.

Our final step in assessing the stability of the phase chimera state (within the subspace $E_4$) is to employ the Routh-Hurwitz criterion, which gives a systematic way of enforcing that all eigenvalues lie in the left half-plane (see, e.g., [3]). For a fourth order system like (12), there are four overlapping constraints, but only two are needed to enforce stability in the real domain where $K \geq 0$ and $C_1, C_2 \in \mathbb{R}$. These reduce to

$$K < \frac{1}{2} \tag{15}$$

and

$$AK^4 + BK^3 + CK^2 + DK + E > 0 \ , \tag{16}$$

where

$$A = -2C_1^4 - 8C_2C_1^3 + [(-9\delta^2+1)C_2^2 - 9\delta^2 - 27]C_1^2 - 72C_1C_2 + (-9\delta^2+9)C_2^2 - 9\delta^2 - 81 \ ,$$
$$B = C_1^4 + 12C_2C_1^3 + [(21\delta^2-1)C_2^2 + 21\delta^2 + 45]C_1^2 + 204C_1C_2 + (21\delta^2-21)C_2^2 + 21\delta^2 + 216 \ ,$$
$$C = -4C_2C_1^3 + (-16\delta^2C_2^2 - 16\delta^2 - 24)C_1^2 - 212C_1C_2 + (-16\delta^2+16)C_2^2 - 16\delta^2 - 216 \ ,$$
$$D = [(4\delta^2C_2^2 + 4\delta^2 + 4)C_1^2 + 96C_1C_2 + (4\delta^2-4)C_2^2 + 4\delta^2 + 96 \ ,$$
$$E = -16C_1C_2 - 16 \ .$$

The main text gives the simplified version of this complicated condition for $C_2 = -C_1$ and also gives the limits (in that case) for $x \to 0$ and $x \to 1/2$.

## HISTORICAL NOTES

In the 1990s, Hakim, Rappel, and Chabanol performed calculations to determine the stability of the incoherent case (Region I below the dotted curve in Fig. 1 of Sethia and Sen's 2014 paper [4]) and in fact, performed stability calculations in an even more general case for arbitrary values of $\delta$ (the stability region in the $\delta \to 1$ limit corresponds to the dot-dashed curve in Fig. 1 of Sethia and Sen's 2014 paper). In 1992, Hakim and Rappel [5] observed:

> "However, states distributed unevenly on the limit circle (but still with A = 0) also exist and have a slightly larger domain of stability."

Five years later, Chabanol, Hakim, and Rappel noted that "There are of course many different ways to realize condition (2)." What they did not notice was that one of the ways of satisfying the condition is actually the phase chimera state (and multitailed phase chimera states). Indeed, chimera states were not even given a name until 2004 [6].

In a 2013 paper, Sethia, Sen, and Johnston [7] reported their discovery of the Amplitude Mediated Chimera (AMC) in a nonlocally coupled version of the Stuart-Landau equation. Their numerical explorations showed a version of the phase chimera (which they refer to as a "phase-only chimera state" in the paper) in the the weak coupling limit with a small value for the parameter $K$, in the same nonlocally coupled version of the Stuart-Landau equation. Beyond the fact that it was in a nonlocal coupling version of the Stuart-Landau equation, the difference with the phase chimera we report here is that the long-term average frequencies are not uniform (as is the case here) and the order parameter is not equal to 0. Just three months after this first paper in 2013, Sethia and Sen reported the existence of the same AMC in a globally coupled system with linear coupling [4], but they make no mention of the "phase-only chimera"

from their earlier paper. Today, nine years later, we report the discovery in the globally coupled Stuart-Landau equation of the phase chimera, a fascinating pattern of disorder frozen in the co-rotating frame.

---

* emmazajdela@u.northwestern.edu
† dmabrams@northwestern.edu



\end{document}